\documentclass[twocolumn,twoside,slac_two]{revtex4}
\usepackage{graphicx}
\usepackage{fancyhdr}
\usepackage{graphics}
\usepackage{epstopdf}
\usepackage{textpos}
\usepackage{hepunits2007}
\usepackage{terascale_defs}
\begin{document}

\title{\centering Determinations of the Strong Coupling at HERA}
\author{
\centering
\begin{center}
T. Sch\"orner-Sadenius (on behalf of the H1 and ZEUS Collaborations)
\end{center}}
\affiliation{\centering Deutsches Elektronen-Synchrotron (DESY), Notkestr. 85, 22607 Hamburg, Germany}
\begin{abstract}
The status of determinations of the QCD coupling constant, \alpS, at HERA is reviewed. 
Since jet final states provide the most relevant
input to the HERA determinations of \alpS, the relevant methods used in and results from jet physics are also discussed. Furthermore, HERA and world averages of \alpS values are presented. Finally, the HERA-PDF 1.6 proton parton distribution function set which also uses jet final-state data is introduced. 
\end{abstract}

\maketitle
\thispagestyle{fancy}

\section{\label{sec:intro}INTRODUCTION}
The world's only \ep collider, HERA at DESY, Hamburg, concluded operations in 2007 after 15 years of data taking. Until then, integrated luminosities of about \unit{0.5}{\invfb} had been collected by each of the two colliding-beam experiments H1 and ZEUS. An integral part of the HERA physics programme are studies of the hadronic final state, often in the shape of hard hadronic jets, with the aim of testing predictions of (perturbative) QCD (pQCD). Consequently, much effort has gone into extractions of the strong coupling constant, \alpS, from jet measurements.\\
Values of \alpS have also been extracted from inclusive measurements which are sensitive to \alpS and to the gluon density in the proton, $g$, only via scaling violations. Furthermore, in these measurements $g$ and \alpS are strongly correlated. The corresponding results on \alpS from the HERA kinematic regime are therefore less precise than those from jets. One way out of this problem is to include jet data in  fits of the proton \emph{parton distribution functions} (PDFs), thus  introducing leading-order sensitivity to $g$ and \alpS and adding quark-induced contributions to the cross section which break the strong $g$--\alpS correlation mentioned above. A recent PDF fit result makes use of this feature (HERA-PDF 1.6). \\
This article reviews measurements of the strong coupling constant at HERA, with an emphasis on the relevant jet measurements, on combinations of \alpS values and on combined extractions of the PDFs and \alpS.

\section{\label{sec:environment}EXPERIMENTAL ENVIRONMENT}
The HERA \ep collider\footnote{
HERA could accelerate both electrons and positrons. Since the lepton charge sign is of no importance for the physics  discussed here, the term ``electron'' will be used generically for both electrons and positrons. 
} was in operation from 1992--2007. The electron beam energy was fixed to \unit{27.5}{\GeV}. 
The proton beam energy was raised from \unit{820 to 920}{\GeV} during the ``HERA-I'' data-taking phase (1998) which lasted until 2000. With these beam energies, HERA achieved a centre-of-mass energy of about \unit{318}{\GeV}.   
A shutdown in the years 2001--2003 was used to prepare the HERA-II phase which brought an increase in luminosity by a factor 4--5 and the possibility of longitudinal lepton polarisation.\\
The HERA experiments H1 and ZEUS were typical high-energy physics detectors. The most striking feature of both detectors was their asymmetric structure which reflected the different beam energies of proton and lepton beam and the boosted centre-of-mass system. Both H1 and ZEUS provided tracking with silicon detectors close to the interaction point and large-volume jet drift chambers. Typical \pT resolutions were of the order of $\sigma\left(\pT\right) / \pT = 0.003 \cdot \pT \left[\GeV\right]$ for both experiments, with an additional constant contribution of $0.015$ in the case of H1. The tracking devices were surrounded by calorimeters (LAr sampling technology with 45\,000 cells and lead / steel absorbers in H1, depleted uranium with scintillator for the compensating ZEUS calorimeter with about 12\,000 cells). The energy resolutions for electrons were determined in test beam measurements to be 11(18)$\% / \sqrt{E \left[ \GeV \right]}$ for electrons in the case of H1 (ZEUS). For hadrons, the ZEUS performance was about 
35$\% / \sqrt{E \left[ \GeV \right]}$, for H1 50$\% / \sqrt{E \left[ \GeV \right]}$  were achieved. \\
In both experiments, superconducting coils provided magnetic fields of \unit{1.16(1.43)}{\tesla} (H1 / ZEUS). The experiments were additionally equipped with small-angle detectors for the measurement of particles close to the beam pipe, and with large-area muon chambers for the rejection of cosmic-ray background and the measurement of muons and hadronic leakage. Luminosity detectors which employed the Bethe-Heitler (or \emph{bremsstrahlung}) process \ep$\rightarrow$\ep$\gamma$ complemented the detectors.  

\section{\label{sec:physics}QCD AND PHYSICS AT HERA}
\subsection{\label{sec:physics:qcd}QCD and the Strong Coupling}

The strong coupling constant \alpS is the central parameter of QCD. It obeys the \emph{renormalisation group equation} (RGE)
\begin{equation}
\qtwo \frac{ \partial \alpS \left( \qtwo \right) }{\partial \qtwo} = \beta\left( \alpS \left( \qtwo\right)\right)\, .
\end{equation}
The $\beta$ function can be calculated using perturbative methods; so far contributions up to four-loop precision have been derived. The solution for \alpS at the two-loop level (which is the relevant one for usage in NLO pQCD calculations of jet cross sections) is  given by 
\begin{equation}
\alpS\left(\qtwo\right) = \frac{1}{\beta_0 L} - \frac{1}{\beta_0^3 L\squared}\beta_1 \ln L\, ,
\end{equation}
with known coefficients $\beta_{0,1}$ and $L = \ln \left( \qtwo / \Lambda\squared \right)$. $\Lambda$ defines the energy scale at which the perturbative approach breaks down. Its value has to be determined from data and is of the order of \unit{200}{\MeV}. 

The RGE fixes the behaviour of \alpS with energy scale but not the absolute normalisation which has to be taken from data. Alternatively, one can fix the coupling value for a given reference energy scale $\mu$ and express the value at other scales \qtwo as a function of the reference scale. For the one-loop solution this leads to 
\begin{equation}
\alpS\left( \qtwo \right) = \frac{\alpS\left( \mu\squared \right)}{1+\alpS\left( \mu\squared\right) \beta_0 \ln \frac{\qtwo}{\mu\squared}}\, .
\label{eq:evolution}
\end{equation}
Typically the reference scale $\mu$ is chosen to be the mass of the \Zo boson, \MZ. Equation~\ref{eq:evolution} can be used to evolve the strong coupling from the scale \MZ\,  --- at which it is often measured --- to any arbitrary scale that might be needed in a calculation.

\subsection{\label{sec:physics:kinematics}Basics of HERA Physics}
\begin{figure}
\includegraphics[width=75mm]{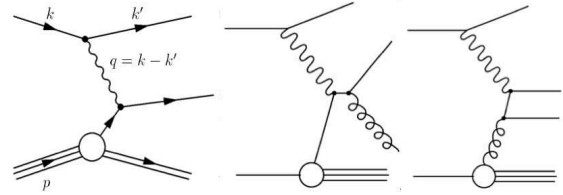}
\caption{Feynman diagrams of important processes in \ep physics.}
\label{fig:feynman}
\end{figure}

Figure~\ref{fig:feynman} (left) shows a Feynman diagram for a lowest-order \ep scattering process. The incoming 
electron with four-momentum $k$ interacts with the proton (momentum $P$) via the exchange of a boson of four-momentum $q$, resulting in a final state with a scattered lepton ($k'$) and a hadronic final state $X$. The kinematics of this process are fully described by the following quantities: The \emph{momentum transfer} $\qtwo = -q\squared = -(k-k')\squared$  defines the resolution power of the exchanged boson. The \emph{Bjorken scaling variable} $x = \qtwo / (2P\cdot q)$ defines the proton's momentum fraction that participates in the hard interaction. The \emph{inelasticity} $y = (P \cdot q)/(P \cdot k)$ is the fractional energy transferred from the electron to the proton in the latter's rest frame. The quantities \qtwo, $x$ and $y$ are related via the squared centre-of-mass energy, $s$, such that for given beam energies two of the three are sufficient to fully characterise the kinematics of the scattering process: $\qtwo = xys$.

Two further distinctions are made: First, the observable \qtwo is used to define two different kinematic regimes: \emph{photoproduction} (PHP) is defined by the condition $\qtwo \sim 0$; for \qtwo significantly larger than \unit{1}{\GeV\squared} one speaks of \emph{deep-inelastic scattering}.  Second, the type of the exchanged boson is used to distinguish between \emph{neutral-current} (NC, for electron / \Zo-boson exchange) and \emph{charged-current} (CC, \Wpm-boson exchange) events. In the latter case, the final-state lepton is a neutrino and will escape the experiment undetected, leading to missing transverse energy in the final state. For determinations of the strong coupling constant at HERA, CC jet physics is of negligible importance. 

The kinematics of the \ep scattering are typically derived from measurements of the scattered electron or, in the case of CC events or of photoproduction events (where the electron escapes the experiment undetected), from the properties of the hadronic final state.  

The hadronic final state was originally calculated from the energy depositions in the calorimeters. Later on, \emph{energy-flow algorithms} were developed in both experiments which aim at maximising the resolution by making the best use of both tracking and calorimetric information. Also jets were either reconstructed from calorimeter cells (mostly for ZEUS) or from the energy-flow objects (typically in H1). The hadronic energy scale is known, in both experiments, to about 1\%, and this uncertainty is in most cases the dominating experimental one, followed often by the model uncertainty that is evaluated by using different MC models for the correction of the data for detector effects. 

The experiments at HERA typically measure the so-called \emph{reduced cross section}, $\sigma_r$, that is closely related to the double-differential cross section in the kinematic quantities \qtwo and $x$ and that is to a good approximation given by the structure function $F_2$:  
\begin{equation}
\sigma_r = \frac{1}{Y_+}\frac{\dif\squared\sigma_{NC}^{\pm}}{\dif x \dif \qtwo}\frac{xQ^4}{2\pi\alpha\squared}=F_2\left(1+\Delta\right)\, ,
\end{equation} 
with $Y_{\pm} = 1\pm\left(1-y\right)\squared$.

In leading order, $F_2$ is related to the PDFs --- or more precisely the quark densities, $q_i\left(x,\qtwo\right)$ --- via the relation $F_2 = x \sum_i e\squared_i \left(q_i + \overline{q}_i\right)$, where the sum runs over all quark flavours and $e_i$ is the charge of the quark $i$ in units of the elementary charge. From a measurement of $F_2$, therefore, the PDFs can be extracted using the DGLAP equations which govern the evolution of the PDFs with \qtwo. 

At HERA, the HERA-PDF working group of the H1 and ZEUS experiments has since a few years extracted PDF sets from combined H1+ZEUS reduced cross section data. The latest HERA-PDF set based entirely on inclusive (structure-function) data is known as HERA-PDF 1.5~\cite{ringaile}. 
 
\subsection{\label{sec:physics:jets}Jet Physics at HERA}

The inclusive measurements sketched above rely exclusively on the measurement of the scattered electron and the subsequent  determination of the kinematics. Neither is the hadronic final state involved, nor does the leading-order contribution (left diagram in \fig{\ref{fig:feynman}}) involve a QCD coupling --- the process is of a purely electroweak nature ---, nor do the inclusive data provide a direct access to the gluon density in the proton, $g$ (an indirect access that is also strongly correlated with \alpS is given via the scaling violations of the DGLAP evolution equations).
All these limitations for more extensive studies of QCD can be overcome by studies of jets. 

Jets are (collimated) bundles of hadrons which are created by the showering and hadronisation of final-state partons. Since the involved processes do not lead to significant transverse momenta, the resulting hadrons are close in phase space. A jet is formed from these hadrons by applying a specific procedure or \emph{jet algorithm} that defines which particles are combined into a jet and how the resulting jet four-momentum is calculated from the four-vectors of the contributing particles. The limited transverse momenta also ensure that the jet's four-momentum is close to the original parton's so that the jets can be regarded as the ``footprints'' of the final-state partons and can give direct access to the hard interaction. At HERA, typically the \emph{longitudinally invariant inclusive \kT algorithm}~\cite{inclkt} is used. For DIS analyses, the algorithm is typically applied in the \emph{Breit reference frame} in which the transverse momenta of the jets are a measure for the hardness of the underlying QCD process. 

The two Feynman diagrams in the centre and on the right side of \fig{\ref{fig:feynman}} show the dominant contributions to jet production at HERA; they are of order $\mathcal{O}\left(\alpS\right)$ (centre: \emph{QCD-Compton process} or QCDC; right: \emph{boson-gluon fusion} or BGF). The BGF process introduces a dependence of the jet production cross section on the gluon density, $g$, already at leading order. The QCDC process, on the other hand, helps to break the aforementioned strong correlation between the gluon density and \alpS from which inclusive measurements suffer.

The cross section for jet production can be written as a convolution of the proton PDFs $f_{i/p}$ with a hard scattering matrix element $\hat{\sigma}$, expanded in powers of \alpS: 
\begin{eqnarray}
\sigma_{\text{jet}} & = & \sum_n \alpS^n\left(\murs\squared\right) \\
& & \cdot  \sum_{i=q,\overline{q},g} \int \dif x f_{i/p} \left(x,\mufs\squared\right) \cdot C_{i,n}\left(x,\murs\squared,\mufs\squared \right)\, .\nonumber
\end{eqnarray} 
In this equation, the $f_i$ are the parton distribution functions, and the coefficients $C_i$ can be calculated --- up to some order --- in pQCD. The \murs and \mufs are the renormalisation and factorisation scales. The terms proportional to $\alpha_s^2$ or higher correspond to corrections to the leading-order diagram. 
In pQCD, jet cross sections can be calculated up to next-to-leading order (NLO) for the case of inclusive-jet, dijet and trijet production (DIS) or dijet production (PHP). Measurements of jet cross sections have been performed for all of these scenarios in various regions of phase space (different regions of \qtwo, various requirements on transverse energies or pseudorapidities of the jets, etc.). The dominating theoretical uncertainty (and often the largest uncertainty overall) is typically the effect of unknown higher orders (beyond NLO) in the perturbative expansion. This uncertainty is typically estimated by a variation of the renormalisation scale, \murs, in the calculations. 

\section{\label{sec:alphas}THE STRONG COUPLING AT HERA}

\subsection{\label{sec:alphas:strong}Deriving the Strong Coupling at HERA}
Two different methods for deriving the strong coupling from HERA jet data exist. ZEUS~\cite{zeusalphasfit} typically performs NLO QCD calculations with PDF sets that use different input \alpS values, making it possible to parametrise the dependence of the theory cross section in a given analysis bin $i$ of an observable $A$ on \alpS by a quadratic function:
\begin{equation}
\frac{\dif\sigma}{\dif A}\bigg\vert_i = C_1 \cdot \alpS\left(\MZ\right) + C_2 \cdot \alpS\squared\left(\MZ\right)\, .
\end{equation}
The cross section measured in bin $i$ is then mapped to the parametrisation and the value of \alpS can easily be read of. In this way, the full coupling dependence of the PDFs and the hard scattering matrix element is preserved, and experimental and theoretical uncertainties can easily be derived. The method is applicable to both single data points and to sets of several points.  

H1, on the other hand, employs the Hessian method~\cite{hessian} for deriving values of \alpS.  In this method a full $\chi\squared$ is evaluated for an arbitrary number of data points: 
\begin{equation}
\chi\squared = \vec{V}^T \cdot M^{-1} \cdot \vec{V} + \sum_k \epsilon\squared_k\, ,
\label{eq:chi2}
\end{equation}
where the matrix $M$ comprises the statistical and uncorrelated systematic uncertainties and $V_i = \sigma_i^{exp} -  \sigma_i^{theo} \left( 1- \sum_k\Delta_{ik}\epsilon_k\right)$. The $\sigma_i$ are the measured and theoretically predicted cross sections for bin $i$, $\Delta_{ik}$ is the correlated systematic uncertainty of type $k$ for bin $i$, and $\epsilon_i$ is a fit parameter which in \eqn{\eqref{eq:chi2}} is used as a penalty term.  The experimental uncertainty can be read of at the places where $\chi\squared = \chi\squared_{min} + 1$. 
The Hessian method is also used in the combined HERA \alpS determinations discussed below. 

\subsection{\label{sec:alphas:jetsphp}\boldmath{\alpS} from PHP Jet Data}
\begin{figure}
\includegraphics[width=65mm]{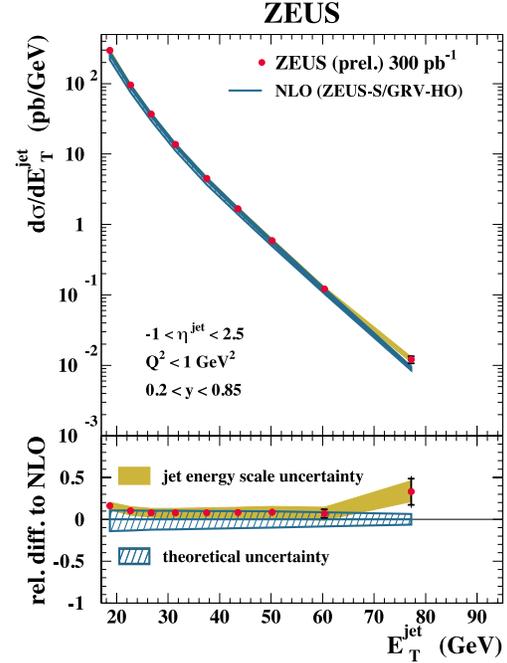}
\caption{ZEUS inclusive-jet cross section in PHP compared to NLO predictions.}
\label{fig:zeusphp}
\end{figure}

ZEUS recently released a preliminary determination of  \alpS from a PHP measurement of inclusive-jet cross sections in \unit{300}{\invpb} of HERA-II data~\cite{zeusphppaper}. The result of the measurement is shown in \fig{\ref{fig:zeusphp}} as a function of the transverse jet energy, \ET. The measurement is experimentally limited by the jet energy scale uncertainty which amounts to 1.8\% on the extracted values of $\alpS\left(\MZ\right)$; the theoretical uncertainties are dominated (as is the case for most HERA jet measurements) by the influence of missing higher orders in the perturbative expansion of the cross section. The effect on \alpS is estimated to be of the order of 2.5\%. The resulting value for \alpS is 
\begin{equation*}
\alpS\left(\MZ\right) = 0.1206^{+0.0023}_{-0.0022}\left(exp\right)^{+0.0042}_{-0.0033}\left(theo\right)\, .
\end{equation*}
A similar measurement~\cite{zeusphppaperold} in slightly smaller statistics (\unit{189}{\invpb}) has also been carried out using the anti-\kT and \SISCONE jet algorithms (instead of only the \kT algorithm) which are by now the default choices at the LHC. Compatible values of the strong coupling have been extracted from all three measurements. 
  
\subsection{\label{sec:alphas:jetsdis}\boldmath{\alpS} from DIS Jet Data}

In PHP jet measurements, the transverse jet energy, \ET, is used as a hard scale. In DIS, also the photon virtuality \qtwo can be used (or linear combinations of \ET and \qtwo). However, at low values of \qtwo, the theoretical uncertainty due to the effect of missing higher orders in the perturbative expansion is typically significant. 

H1 measured inclusive-jet, dijet and trijet cross sections at low values of \qtwo between \unit{5 and 100}{\GeV\squared} in HERA-I data~\cite{h1lowq2paper}. The measurements were performed (multi-)differentially as functions of e.g.\ \qtwo and jet \pT. As a first step, values of \alpS have been extracted from all 61 measured data points individually. The individual values are observed to be well compatible.

In a next step, combinations of all data points in 4 different \qtwo intervals were fitted. Since the experimental uncertainties of the resulting \alpS values are typically much smaller than the theoretical ones, the results demonstrate the limited predictive power of NLO calculations at least at low \qtwo values and thus the necessity of going to higher orders in the perturbative expansion, i.e.\ of using calculations at next-to-next-to-leading order (NNLO). 

Finally, all 61 data points are included in one combined fit, resulting in an \alpS value of  
\begin{eqnarray*}
\alpS\left(\MZ\right) &=& 0.1160\pm0.0014\left(exp\right)\\
& &^{+0.0093}_{-0.0077}\left(theo\right)\pm0.0016\left(PDF\right)\, .
\end{eqnarray*}

The di- and trijet cross sections just discussed were also used by the H1 collaboration for a measurement of the strong coupling from the ratio $R_{3/2}$ of trijet to dijet cross section. A possible benefit of this ratio is a (partial) cancellation of some systematic uncertainties between numerator and denominator (luminosity, scales, PDFs ...). However, mainly because of limited statistics the result alone is not competitive with the best determinations from e.g.\ inclusive-jet cross sections at high \qtwo (see later).  A similar analysis has also been performed by ZEUS~\cite{zeuslowq2paper}.

\subsection{\label{sec:alphas:jetshighq2}\boldmath{\alpS} at the Highest \boldmath{\qtwo} Values}

\begin{figure}
\includegraphics[width=60mm]{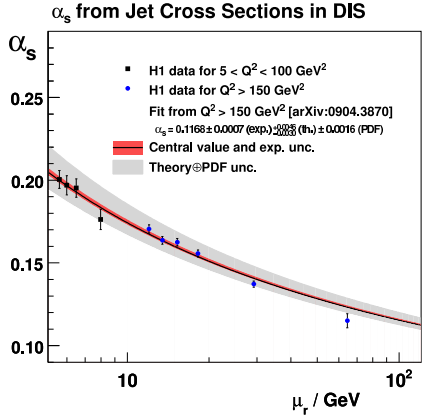}
\caption{H1 \alpS values from combinations of 1/2/3-jet cross sections in high- and low-\qtwo measurements.}
\label{fig:h1highq2a}
\end{figure}

A 2009 publication from H1 complements the above-mentioned low-\qtwo measurement of inclusive-, di- and trijets by the corresponding measurements in the high-\qtwo regime (above \unit{150}{\GeV\squared})~\cite{h1highq2paper1}. The extracted value is $\alpS = 0.1168\pm 0.0007(exp.)^{+0.0046}_{-0.0030}(th.)\pm 0.0016(PDF)$. 
Figure~\ref{fig:h1highq2a} shows the main result, namely the running of the coupling as derived from a combined fit to all high-\qtwo data points (solid line with error band), together with individual \alpS values (evolved to their respective scale) at high \qtwo (small circles). 
In the figure, the high-\qtwo value is extrapolated down to the low-\qtwo regime of the measurement described above. The good agreement of the extrapolation with the four low-\qtwo measurements (small squares) is as striking as is the strong reduction of the theoretical uncertainties with respect to the low-\qtwo \alpS result.

It is due to this smaller sensitivity to theoretical uncertainties that most \alpS measurements at HERA have been performed in the region of high \qtwo values. The most recent published ZEUS and H1 results comprise a measurement of inclusive-jet cross sections at $\qtwo > \unit{125}{\GeV\squared}$ with the \kT, anti-\kT and \SISCONE jet algorithms in \unit{82}{\invpb} of HERA-I ZEUS data, and a multi-differential measurement measurement of inclusive/di/tri-jet cross sections in \unit{351}{\invpb}  at $\qtwo > \unit{150}{\GeV\squared}$ from H1. 

The ZEUS measurement~\cite{zeushighq2paper1} used the $\dif \sigma / \dif \qtwo$ distributions at \qtwo values above \unit{500}{\GeV\squared} for the extraction of the strong coupling, since this region proved to have the smallest theoretical uncertainties. 
The agreement  observed between data and NLO QCD is excellent, and up to \qtwo values of about \unit{1000}{\GeV\squared} the theory uncertainty is found to dominate the measurement. 
ZEUS extracted a value of the strong coupling for each of the three jet algorithms; the three values are very well compatible, and the \kT result is given as 
\begin{eqnarray*}
\alpS\left(\MZ\right) &=& 0.1207\pm0.0014\left(stat\right)\\
&&^{+0.0035}_{-0.0033}\left(exp\right)^{+0.0022}_{-0.0023}\left(theo\right)\, .
\end{eqnarray*} 
The analysis has also been repeated in about \unit{300}{\invpb} of HERA-II data~\cite{zeushighq2paper2}, and the two results agree very well.
  
\begin{figure}
\includegraphics[width=75mm]{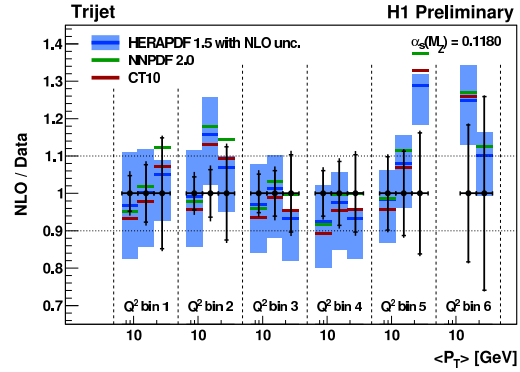}
\caption{H1 ratio of trijet cross sections to NLO predictions in different regions of \qtwo as functions of the average jet \pT.}
\label{fig:h1newhighq2trijetsalphas}
\end{figure}

The latest H1 high-\qtwo measurement~\cite{h1highq2paper2} was the first H1 jet measurement to profit from a strongly improved calibration of the hadronic final state and jet energy scale. In addition, the statistics of \unit{351}{\invpb} were large enough to allow the first double-differential measurement of trijet cross sections to be performed. As in the low-\qtwo case, H1 extracted \alpS values from each measured data point of the (average) \pT distributions for inclusive jets, dijets and trijets in different regions of \qtwo. The individual measurements agree very nicely as can be seen in \fig{\ref{fig:h1newhighq2trijetsalphas}} which shows the resulting \alpS values for the trijet cross sections for different average jet \pT values and \qtwo regions. Typically, the scale uncertainties are larger than the experimental uncertainties by a factor of about 2. 

In a next step, again, \alpS values are extracted from a combined fit to all data points within a certain \qtwo region, and from all combined inclusive-/di/tri-jet data points. As it turns out, the value extracted from the trijet data points offers the smallest experimental uncertainty. This is due to the fact that the trijet cross section already at leading order is proportional to $\alpha_s^2$.  The resulting value is 
\begin{eqnarray*}
\alpS\left(\MZ\right) &=& 0.1196\pm0.0016\left(exp\right)\\
&&\pm0.0010\left(PDF\right)^{+0.0055}_{-0.0039}\left(theo\right)\, ;
\end{eqnarray*} 
the inclusive-jet and dijet results are compatible.

\subsection{\label{sec:alphas:shape}Alternatives: \boldmath{\alpS} from Jet Shapes}
Jet final states offer still more accesses to the strong coupling value than only via the cross-section measurements discussed above. One example is a ZEUS extraction of \alpS from the averaged integrated jet shape, $\langle \psi\left( r \right) \rangle$, in DIS~\cite{DESY04072}. The quantity $\langle \psi\left( r \right) \rangle$ is defined as $\langle \psi\left( r \right) \rangle = \frac{1}{N_{jets}}\sum_{jets} \frac{\ET\left( r \right)}{\ETjet}$, where $N_{jets}$ is the number of jets studied, $\ET\left( r \right)$ is the transverse energy contained in a cone of radius $r$ ($0 < r < R$, with the jet radius $R$) around the jet axis, and \ETjet is the total jet transverse energy.  $\langle \psi\left( r \right) \rangle$ is thus a measure for the distribution of transverse energy inside a jet, and an understanding of this quantity gives interesting insights into the fragmentation process and allows, among other things, gluon and quark jets to be discriminated against each other on a statistical basis.

\begin{figure}
\includegraphics[width=65mm]{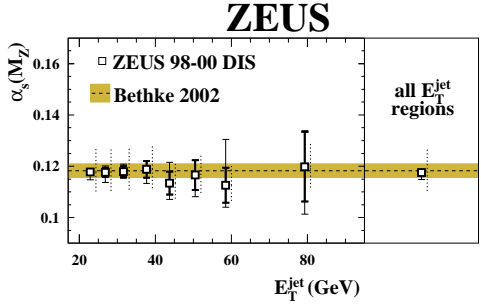}
\caption{ZEUS \alpS values extracted from integrated jet shapes at radius 0.5 as a function of \ETjet. Dashed error bars: theory uncertainties; inner/outer solid error bars: statistical / full experimental uncertainties.}
\label{fig:shapes}
\end{figure}

For the extraction of \alpS, ZEUS parametrised the cross section for  $\langle \psi\left( r \right) \rangle$ at a radius $r = 0.5$ in different bins of \ETjet as  $\langle \psi\left( r \right) \rangle = C_1 + C_2 \cdot \alpS\left( \MZ \right)$. The constants $C_i$ were taken from a fit to the theoretically predicted cross section at different values of $\alpS\left(\MZ\right)$. Since the predictions include only terms of order $\mathcal{O}\left( \alpS \right)$, a reduced sensitivity to \alpS is obtained:  
\begin{eqnarray*}
\alpS\left(\MZ\right) &=& 0.1176\pm0.0009\left(stat\right)\\
&&^{+0.0009}_{-0.0026}\left(exp\right)^{+0.0091}_{-0.0072}\left(theo\right)\, .
\end{eqnarray*} 
Figure~\ref{fig:shapes} shows the \alpS values extracted for the individual \ETjet bins and for the combined extraction. 

\section{\label{sec:heracomb}THE HERA \boldmath{\alpS} COMBINATIONS}

Since 2004, the HERA collaborations have started to combine their \alpS measurements~\cite{heracomb2004}. Since 2007, combined fits to H1 and ZEUS data sets have been performed. The 2007 HERA average, which was based on an inclusive-jet $\dif \sigma/\dif \qtwo$ measurement from ZEUS and the double-differential inclusive-jet $\dif\squared \sigma / \dif \ET \dif \qtwo$ from H1, resulted in~\cite{heracomb2007}
\begin{equation*}
\alpS\left(\MZ\right) = 0.1198\pm0.0019\left(exp\right)\pm0.0026\left(theo\right)\, .
\end{equation*} 
Again, the result is dominated by the theory uncertainties, underlining the necessity of higher orders in the jet cross-section calculations. 

All in all, the HERA combined \alpS measurements demonstrate the excellent agreement between results obtained from various jet final states, from different kinematic regimes (DIS and PHP), and with different methods (H1 and ZEUS). Furthermore, the running character of the coupling can be demonstrated with data from one experimental environment alone --- and it is in excellent agreement with the running as predicted by QCD based on the world average of \alpS (see also later).  

\section{\label{sec:pdf}COMBINED FITS OF PDFS AND \boldmath{\alpS}}

Fits to the inclusive structure-function measurements alone provide good insight into the proton structure. However, there are regions in phase space which are not sufficiently covered by structure-function data to meaningfully constrain the PDFs. One example is the gluon density at large values of $x$. Here, the use of jet data might help: First, the jet data provide sensitivity to $g$ already at leading order, and specifically at large $x$, and second, the QCDC contributions to the jet cross sections break the strong correlation between gluon density and \alpS, allowing a meaningful determination of both parameters to be done simultaneously. 

This insight had early impacts e.g.\ in a 2005 ZEUS publication~\cite{desy05050} in which both structure-function data and DIS and PHP jet cross sections were used in PDF fits, leading to the ZEUS-JETS PDF set. The result was a competitive determination of \alpS and a reduction of the uncertainty on the gluon density at medium and large values of $x$ of up to 35\%.

\begin{figure}
\includegraphics[width=65mm]{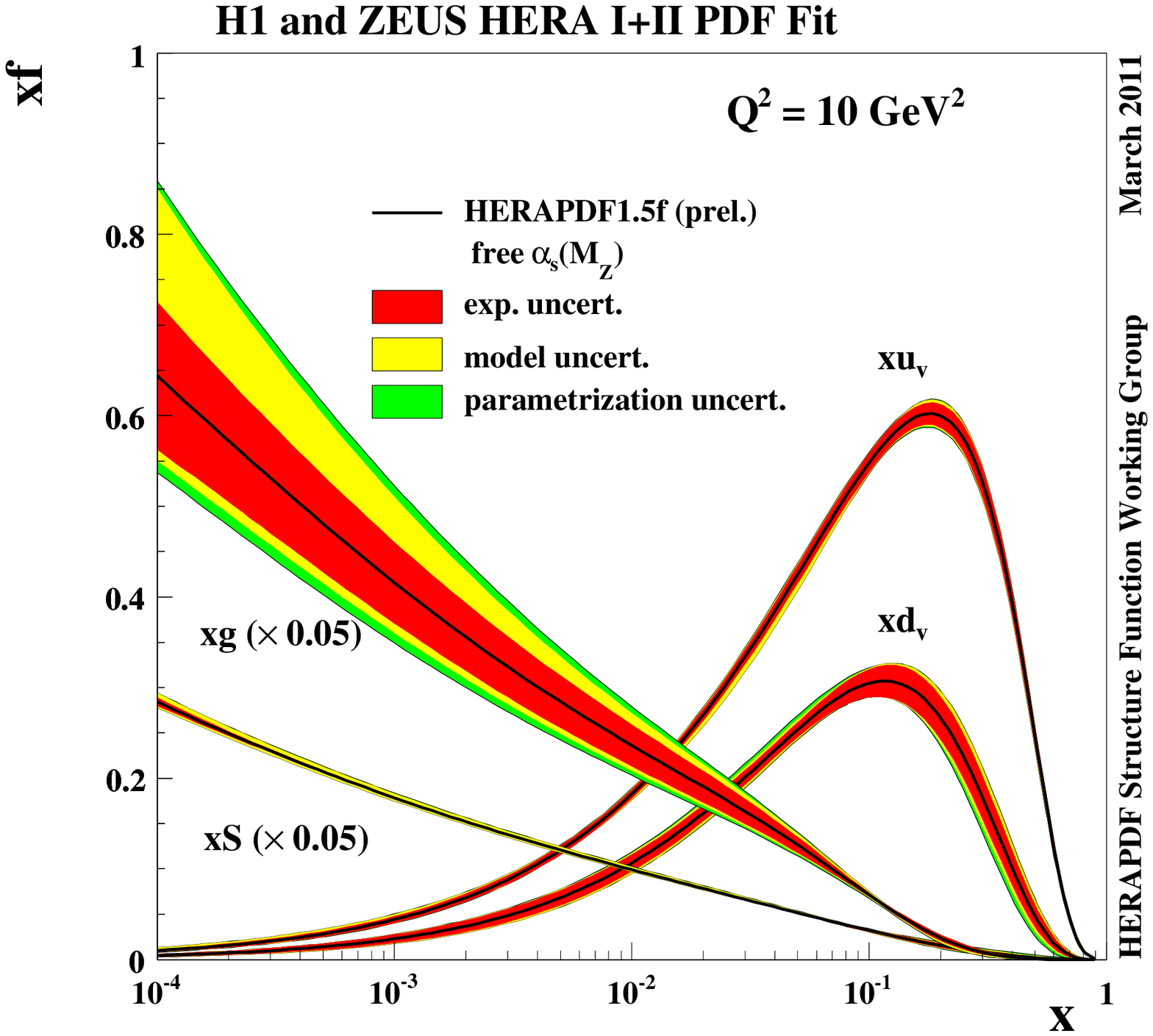}
\includegraphics[width=65mm]{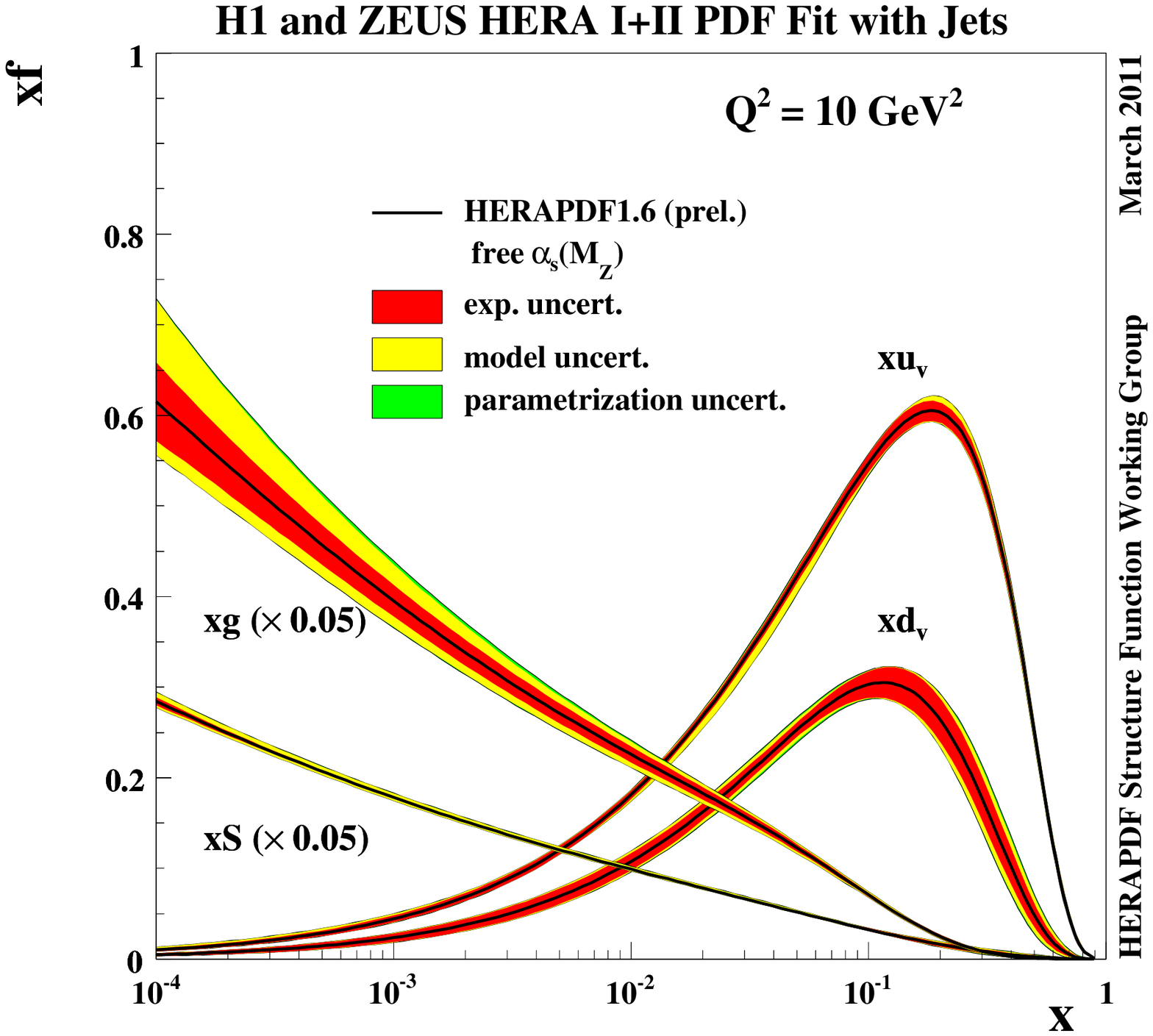}
\caption{Comparison of HERA-PDF 1.5 for free \alpS (top) and 1.6 (bottom).}
\label{fig:herapdf1.6}
\end{figure}

This idea was recently taken up by the HERA-PDF working group who also included jet data from H1 and ZEUS into their PDF fits, leading to the HERA-PDF 1.6 set~\cite{herapdf16}. Detailed studies with fixed \alpS as in HERA-PDF 1.5 showed that the inclusion of the jet data does not change the resulting PDF fit very much, the most remarkable difference being the slightly softer gluon density at high $x$. Freeing \alpS in the fit, however, for the case of HERA-PDF 1.5 suffers  from the already mentioned gluon--\alpS correlation which increases the error on the gluon density. Here HERA-PDF 1.6 (with jets) is clearly superior, with much smaller uncertainties on the gluon density as can be seen from \fig{\ref{fig:herapdf1.6}} which shows, at the top, HERA-PDF 1.5 with free \alpS as a function of $x$ and at the bottom HERA-PDF 1.6 for $\qtwo = 10~\GeV\squared$.

The HERA-PDF 1.6 value for \alpS of  
\begin{eqnarray*}
\alpS\left(\MZ\right) &=& 0.1202\pm0.0013\left(exp\right)\pm0.0007\left(model\right)\\
&&\pm0.0012\left(hadronisation\right)^{+0.0045}_{-0.0036}\left(scale\right)
\end{eqnarray*}
is comparable to the HERA 2007 average.

\section{\label{sec:world}HERA AND {\boldmath $\alpha_S$} WORLD DATA}

\begin{figure}
\includegraphics[width=60mm]{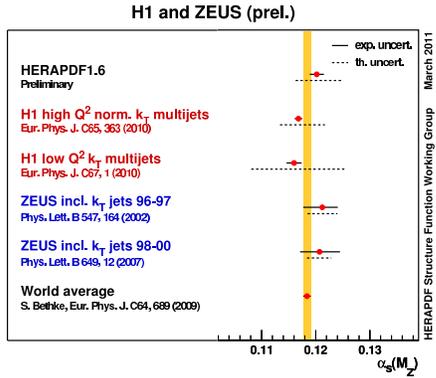}
\caption{Comparison of differnet \alpS values, including the one from HERA-PDF 1.6.}
\label{fig:heraandworld}
\end{figure}

Figure~\ref{fig:heraandworld} shows a comparison of the \alpS value obtained from HERA-PDF 1.6, from the included individual jet measurements from H1 and ZEUS, and the world average~\cite{Bethke:2009jm}. There is excellent agreement, although of course the world average has significantly smaller uncertainties. 

In fact, the precision in the world average value (of order 1\%) comes mostly from the very precise derivations of \alpS from lattice calculations~\cite{davies} and from the analysis of $\tau$ decays. Among the high-energy collider \alpS values contributing to the world average, the determinations from HERA are very competitive.

\section{\label{sec:conclusion}CONCLUSIONS}
HERA has contributed massively to our current understanding of QCD, and in particular to the world knowledge of the proton structure and the central parameter of QCD, \alpS. Values for \alpS have been derived both in analyses of structure functions and from jet measurements in DIS and PHP. Lately, the H1 and ZEUS collaborations have also begun to derive combined values of \alpS from their data sets and to derive values for \alpS in PDF fits which take both inclusive structure-function and jet data into account. The HERA \alpS values are an important input to the \alpS world average.

%

\bigskip 

\end{document}